\documentclass[aps,prb,10pt,superscriptaddress,floatfix,twocolumn,notitlepage]{revtex4-2}
\usepackage{graphicx,graphics}
\usepackage{amsmath,amssymb,amsfonts}
\usepackage{latexsym,verbatim}
\usepackage{bm}
\usepackage{color}
\usepackage{ulem}
\usepackage[percent]{overpic}
\usepackage[breaklinks=true,colorlinks,citecolor=blue,linkcolor=blue,urlcolor=blue]{hyperref}
\usepackage{esint}

\DeclareMathOperator\sgn{sgn}

\DeclareMathOperator\im{Im}
\DeclareMathOperator{\tr}{Tr}
\DeclareMathOperator\sinc{sinc}
\begin{document}
\author{Iacopo Torre}
\email{iacopo.torre@icfo.eu}
\affiliation{ICFO-Institut de Ci\`{e}ncies Fot\`{o}niques, The Barcelona Institute of Science and Technology, Av. Carl Friedrich Gauss 3, 08860 Castelldefels (Barcelona),~Spain}
\author{Lorenzo Orsini}
\affiliation{ICFO-Institut de Ci\`{e}ncies Fot\`{o}niques, The Barcelona Institute of Science and Technology, Av. Carl Friedrich Gauss 3, 08860 Castelldefels (Barcelona),~Spain}
\author{Matteo Ceccanti}
\affiliation{ICFO-Institut de Ci\`{e}ncies Fot\`{o}niques, The Barcelona Institute of Science and Technology, Av. Carl Friedrich Gauss 3, 08860 Castelldefels (Barcelona),~Spain}
\author{Hanan Herzig-Sheinfux}
\affiliation{ICFO-Institut de Ci\`{e}ncies Fot\`{o}niques, The Barcelona Institute of Science and Technology, Av. Carl Friedrich Gauss 3, 08860 Castelldefels (Barcelona),~Spain}
\author{Frank H.L. Koppens}
\affiliation{ICFO-Institut de Ci\`{e}ncies Fot\`{o}niques, The Barcelona Institute of Science and Technology, Av. Carl Friedrich Gauss 3, 08860 Castelldefels (Barcelona),~Spain}
\affiliation{ICREA-Institució Catalana de Recerca i Estudis Avançats, Pg. Lluís Companys 23, 08010 Barcelona, Spain}
\title{Green's functions theory of nanophotonic cavities with hyperbolic materials}
\begin{abstract}
We develop a theory of the quasi-static electrodynamic Green's function of deep subwavelength optical cavities containing an hyperbolic medium. We apply our theory to one-dimensional cavities realized using an hexagonal boron nitride and a patterned metallic substrate.
\end{abstract}
\maketitle
\section{introduction}
\label{sect:introduction}
The rapidly rising field of nanophotonics is based on the confinement of light to length-scales much smaller than the corresponding free-space wavelength $\lambda_{free} = 2\pi /(c\omega)$, where $\omega$ is the angular frequency and $c$ the speed of light.
This allows an enhancement of the light-matter interaction and paves the way to the study of interesting strong-coupling effects.

Confinement of light requires the interaction with a resonant mode of matter to form a polariton.
Different types of polaritons have been used to this aim including surface plasmon polaritons, graphene plasmons, phonon polariton.

Anisotropic materials can have regions of the spectrum, known as {\it reststrahlen} bands, in which the real part of the dielectric permittivity has different signs along different principal axes. 
The prototypical example of a material displaying such a behaviour is hexagonal Boron Nitride (hBN), which is a layered material that has two reststrahlen bands in the mid-infrared region of spectrum, one in the range $95-100~{\rm meV}$, where the out-of-plane permittivity is negative, and one in the range $170-200~{\rm meV}$ where the in-plane permittivity is negative. The two bands lie close to the out-of-plane and in-plane optical phonon resonances respectively.

Within the reststrahlen bands an anisotropic material hosts hyperbolic modes that propagate as waves or collimated beams even in the quasi-static limit.
 
In this work we provide a theory of how hyperbolic polaritons can be confined inside a cavity giving rise to sharp resonances.
The confinement is provided by patterning a nearby metal structure without affecting the quality of the hyperbolic material.
In our work we use hBN as a specific example of hyperbolic material but the same method can equally well applied to any other uniaxial hyperbolic material.

Our theory is based on the calculation of Green functions of few simple building blocks that are then stuck together by imposing appropriate continuity conditions at the contact surface.

The Article is organized as follows. In Section \ref{sect:green} we review the general theory of the Green's function that is relevant for our purpose, in Section \ref{sect:primitive} we calculate the Green's function of the building blocks. In Section \ref{sect:cavities} we use the previous results to calculate the Green's function of a one-dimensional cavity.
In Section \ref{sect:conclusions} we comment and summarize our results. 
Important mathematical details are contained in Appendices. 

\section{Green's theorem}
\label{sect:green}
In the quasi-static limit ($\omega \ll c /L$), where $\omega$ is the angular frequency, $c$ is the speed of light, and $L$ is the characteristic size of the system, Maxwell equations reduce to the equation of electrostatics with the frequency appearing only indirectly through the material permittivities.
This equation, in presence of linear dielectrics, reads (Gauss units are used throughout this work unless explicitly stated)
\begin{equation}\label{eq:poisson}
-\nabla \cdot \left[{\bm \epsilon}(\bm r, \omega) \cdot \nabla \phi(\bm r,\omega) \right] = 4\pi \rho(\bm r, \omega),
\end{equation}
where $\phi(\bm r,\omega)$ is the electric potential, ${\bm \epsilon}(\bm r, \omega)$ is the {\it frequecy dependent} dielectric tensor and $\rho(\bm r, \omega)$ the density of {\it free} charges (i.e. charges not bound in a dielectric).

Once Eq.~(\ref{eq:poisson}) is solved the electric field can be calculated as $\bm E (\bm r, \omega) = - \nabla \phi(\bm r, \omega)$, while the displacement field is given by $\bm D (\bm r, \omega) = -{\bm \epsilon}(\bm r, \omega) \cdot \nabla \phi(\bm r, \omega)$.

The dielectric tensor can be split into its real and imaginary parts as
\begin{equation}\label{eq:dielectric_tensor_splitting}
\bm \epsilon(\bm r, \omega) = \bm \epsilon^{(1)}(\bm r,\omega) + i \bm \epsilon^{(2)}(\bm r,\omega). 
\end{equation}
For real $\omega$, and in absence of time reversal symmetry breaking, the matrices $\epsilon^{(1/2)}(\bm r,\omega)$ are symmetric real matrices.

The uniqueness of the solution of (\ref{eq:poisson}) on a certain region of the space $\Omega$, given Dirichlet or Neumann boundary conditions on the boundary $\partial \Omega$, relies on the positive definiteness of at least one of these two matrices as detailed in Appendix \ref{sect:uniqueness}.

At zero frequency, i.e. in the truly static case, $\epsilon^{(2)}(\bm r,\omega)$ vanishes (since it is frequency/odd) and the uniqueness of the solution is ensured by the positve definiteness of $\epsilon^{(1)}(\bm r,\omega)$.
This is not anymore true at finite frequency in hyperbolic materials, in which $\epsilon^{(1)}(\bm r,\omega)$ has mixed signature.
In this case the presence of a positive defined dissipative component $\epsilon^{(2)}(\bm r,\omega)$, (that can be infinitesimally small) is fundamental to guarantee the uniqueness of the solution of (\ref{eq:poisson}). For this reason we will assume in what follows the presence of such a dissipative component.

Under this assumption, Green's theorem allows to express the general solution of Eq.~(\ref{eq:poisson}) as an integral of a suitable Green's function.
In the case of Dirichlet boundary conditions (potential assigned on the boundary of the domain) this reads
\begin{equation}\label{eq:green_theorem}
\begin{split}
\phi(\bm r, \omega) 
=  & \int_{\Omega} d\bm r' g(\bm r; \bm r', \omega)\rho(\bm r',\omega) +\\
 -\frac{1}{4\pi} & \oint_{\partial \Omega} ds' \phi(\bm r', \omega) \left[\hat{\bm n}(\bm r') \cdot {\bm \epsilon}(\bm r', \omega) \cdot \nabla' g(\bm r; \bm r',\omega)\right].
\end{split}
\end{equation}
Here, $\hat{\bm n}(\bm r')$ is the unit vector normal to the boundary pointing outside the domain $\Omega$ and the Green function $g(\bm r, \bm r',\omega)$ is the solution of
\begin{equation}\label{eq:poisson_green}
-\frac{1}{4\pi}\nabla' \cdot \left[{\bm \epsilon}(\bm r', \omega) \cdot \nabla ' g(\bm r; \bm r',\omega) \right] =  \delta(\bm r-\bm r'),
\end{equation}
with the Dirichlet boundary condition
\begin{equation}\label{eq:dirichlet_green}
g(\bm r;\bm r', \omega) = 0,~\forall \bm r'\in \partial \Omega.
\end{equation}
The proof of Green's theorem for the general case of anisotropic dielectrics and complex permeabilities is reviewed in Appendix \ref{sect:green_proof}.

Since in this work we will consider only linear phenomena we can treat each frequency separately. We will therefore leave the $\omega$ dependence of the Green's function implicit unless needed for the sake of clarity.
Using the Green's identity, the boundary conditions and the {\it symmetry} of the dielectric tensor it can be proven that
\begin{equation}
g(\bm r; \bm r', \omega) = g(\bm r';\bm r,\omega).
\end{equation}
%
\section{Primitive Green's functions}
\label{sect:primitive}
Our calculation of the Green's function of photonic cavities relies on the knowledge of the Dirichlet-Green's function of few simple {\it building blocks} that are then combined together to build the Green's functions of more complex structures.
These building blocks are an uniaxial dielectric slab of thickness $t$ sitting on top of an infinite metallic substrate and a metallic parallelepiped cavity including the limit of one dimension much larger than all the others (line cavity).
In the following we derive the electrostatic Green's functions for these geometries.
\begin{figure}
\centering
\begin{overpic}{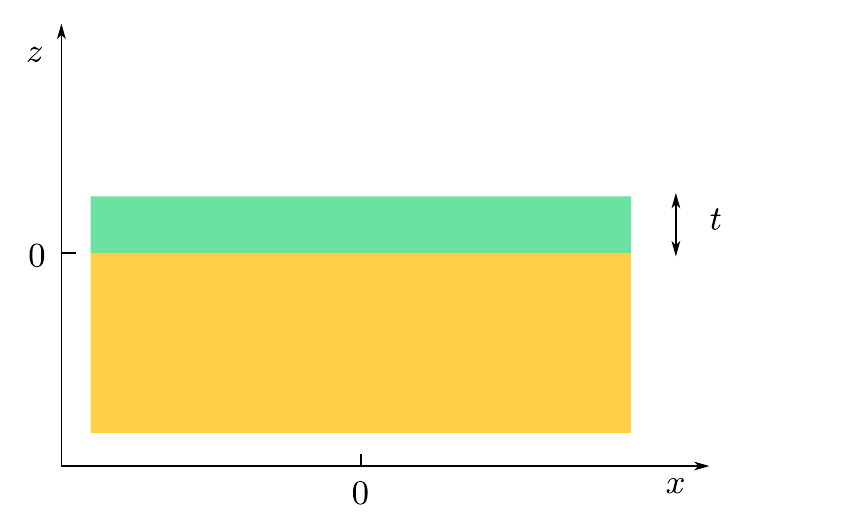}
\put(2, 62){(a)}
\end{overpic}

\begin{overpic}{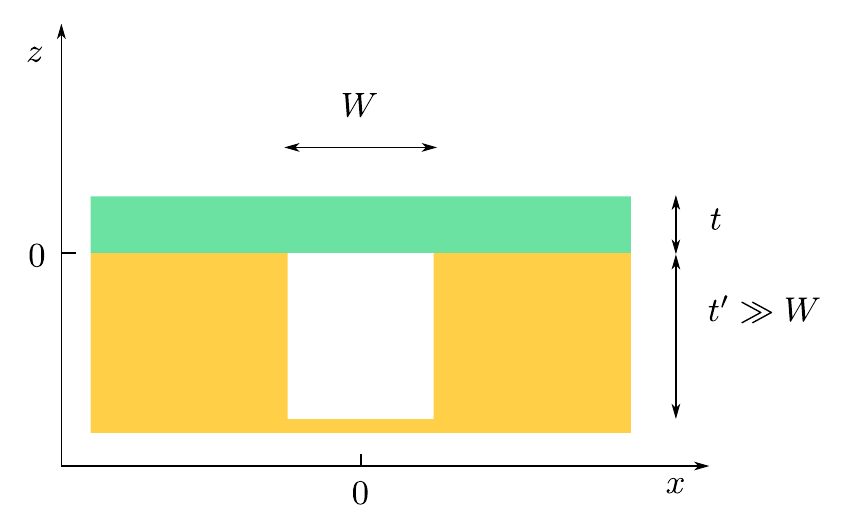}
\put(2, 62){(b)}
\end{overpic}

\caption{\label{fig:geometry}(a) Cross section of an uniaxial dielectric slab on top of a metal. (b) Cross section of the same slab on top of a hollow metallic cavity.  The green rectangle represent an uniaxial dielectric slab, while yellow part represents a perfectly conducting metal.}
\end{figure}
\subsection{Uniaxial dielectric slab on a metal}
\label{sect:slab}
As a first step we calculate the Green's function of a system composed of an uniaxial dielectric slab, with permittivity $\epsilon_\parallel$ in the $x$ and $y$ directions and $\epsilon_\perp$ in the $z$ direction, of thickness $t$ positioned on top of a perfectly conducting metal (See scheme in Fig.~{\ref{fig:geometry}-a).
Since the system is translationally invariant in the $x-y$ directions the solution is more conveniently expressed in Fourier transform with respect to the in-plane direction.
In the following we will denote by an {\it hat} the two-dimensional (2D) in the in-plane directions e.g. $\hat{f}(\bm q) = \int d\bm r e^{-i\bm q \cdot \bm r} f(\bm r)$ and with a {\it tilde} the one-dimensional (1D) Fourier transform along the $y$ direction, i.e. $\tilde{f}(x,q_y) = \int dy e^{-iq_y y} f(x,y)$.

The Green's function $\hat{g}_0^+(z;z';q)$ (note that due to rotational invariance the Green's function only depends on $q = |\bm q|$) solves the Fourier transform of (\ref{eq:poisson_green}) that reads
\begin{equation}\label{eq:poisson_uniaxial}
q^2\epsilon_\parallel(z)\hat{g}_0^+(z;z'; q)-\partial_z[\epsilon_\perp(z)\partial_z \hat{g}_0^+(z;z';q)]=4\pi \delta(z-z').
\end{equation}
Here, the functions $\epsilon_{\parallel(\perp)}(z)$ evaluate to $\epsilon_{\parallel(\perp)}$ for $0<z<t$, and to one for $z>t$.
Equation (\ref{eq:poisson_uniaxial}) must be complemented with the Dirichlet boundary condition at $z=0$, where the metal layer is located
\begin{equation}\label{eq:dirichlet_upper}
\hat{g}_0^+(z=0^+;z';q)=0,
\end{equation}
and with the asymptotic condition $\hat{g}_0^+(z;z';q)\to 0$ for $z \to +\infty$.
It is natural to separate the solution as
\begin{equation}\label{eq:green_function_upper}
\hat{g}_0^+(z;z';q)=\hat{g}^+_{ij}(z;z';q)I_i(z)I_j(z'),
\end{equation}
where $I_1(z)=\Theta(z)\Theta(d-z)$, and $I_2(z)=\Theta(z-d)$, $\Theta(z)$ being the Heaviside step function. Note that $\hat{g}_0^+(z;z';q)$ vanishes if either $z$ or $z'$ is not positive.
The Green's function elements $\hat{g}^+_{ij}$ are given by 
\begin{widetext}
\begin{align}
\hat{g}^+_{11}(z;z';q) & =\frac{2\pi}{q\bar{\epsilon}}e^{-q\eta|z-z'|} -\frac{2\pi}{q\bar{\epsilon}}\frac{(\bar{\epsilon}+1)e^{-q\eta(z+z')}-(\bar{\epsilon}-1)e^{-2q\eta t}\left[e^{q\eta(z+z')}-2\cosh[q\eta(z-z')]\right]}{(\bar{\epsilon}+1)+(\bar{\epsilon}-1)e^{-2q\eta t}},\label{eq:g11}\\
\hat{g}^+_{22}(z;z';q) & =\frac{2\pi}{q}e^{-q|z-z'|}-\frac{2\pi}{q}\frac{e^{q(2t-z-z')}\left[(\bar{\epsilon}-1)+(\bar{\epsilon}+1)e^{-2q\eta t}\right]}{(\bar{\epsilon}+1)+(\bar{\epsilon}-1)e^{-2q\eta t}},\label{eq:g22}
\end{align}
\end{widetext}
and
\begin{equation}\label{eq:g12}
\hat{g}^+_{12}(z;z';q)=\hat{g}^+_{21}(z';z;q)=\frac{8\pi}{q}\frac{e^{-q(z'-t)}e^{-q\eta t}\sinh(q\eta z)}{(\bar{\epsilon}+1)+(\bar{\epsilon}-1)e^{-2q\eta t}},
\end{equation}
were, $\eta=\sqrt{\epsilon_\parallel/\epsilon_\perp }$, and
$\bar{\epsilon}=\eta \epsilon_\perp$.
The real space form of the Green's function can then be obtained, numerically, as
\begin{equation}\label{eq:green_function_upper_real}
\begin{split}
g_0^+(\bm r;\bm r')&  = \int \frac{d \bm q}{(2\pi)^2} e^{i \bm q \cdot (\bm r_\parallel -\bm r_\parallel')}\hat{g}_0^+(z;z';q)\\
& = \int_0^\infty \frac{d qq}{2\pi} J_0(q |\bm r_\parallel -\bm r_\parallel'|)\hat{g}_0^+(z;z';q).
\end{split}
\end{equation}
Here we have separated the in-plane and out-of-plane components of the position vectors as $\bm r = \bm r_\parallel +\hat{\bm z} z$ and $J_0(x)$ is the 0-th order Bessel function of the first kind.
\subsection{Line cavity}
\label{sect:line_cavity}
Here we calculate the Green's function for a metallic hollow cavity located that occupy the region of space with $z<0$ and $-W/2<x<W/2$.
Here we solve for the 1D Fourier transform of the Green function $\tilde{g}^-_0(x,z;x',z';q_y)$ that solves
\begin{equation}\label{eq:poisson_line}
-[\partial_x^2+\partial_z^2-q_y^2]\tilde{g}^-_0(x,z;x',z';q_y) =4\pi \delta(x-x')\delta(z-z'),
\end{equation}
with the boundary condition
\begin{equation}\label{eq:dirichlet_line}
\tilde{g}^-_0(x,z;x',z';q_y) = 0~ \mbox{if} ~z=0, 
\end{equation}
and the asymptotic condition $\tilde{g}^-_0(x,z;x',z';q_y) \to 0$ for $z\to -\infty$.
The solution is found by separation of variables and reads
\begin{widetext}
\begin{equation}\label{eq:green_function_line}
\tilde{g}^-_0(x,z;x',z';q_y) =
2\pi \Theta(-z)\Theta(-z')\sum_{\ell = 1}^\infty\frac{\varphi_\ell(x)\varphi_\ell(x')}{\sqrt{q_y^2 + \frac{\pi^2 \ell^2}{W^2}}}\left[e^{-\sqrt{q_y^2 + \frac{\pi^2 \ell^2}{W^2}}|z-z'|}
-e^{\sqrt{q_y^2 + \frac{\pi^2 \ell^2}{W^2}}(z+z')}
\right].
\end{equation}
\end{widetext}
Here we defined the complete set of orthonormal functions for $n = 1,2, \dots$
\begin{equation}\label{eq:basis_functions_1D}
\varphi_n(x) = I_{\left[-\frac{W}{2},\frac{W}{2}\right]}(x)\sqrt{\frac{2}{W}}\times
\begin{cases}
\sin(n \pi x/W) ~n~\mbox{even}\\
\cos(n \pi x/W) ~n~\mbox{odd}\\
\end{cases},
\end{equation}
where $I_{[-W/2,W/2]}(x)$ is the indicator function of the interval $[-W/2,W/2]$.
The real space form can be recovered from the 1D Fourier transform
\begin{equation}\label{eq:green_function_line_real}
g^-_0(\bm r; \bm r') = \int \frac{dq_y}{2\pi} e^{i q_y (y-y')} \tilde{g}^-_0(x,z;x',z';q_y).  
\end{equation}
\begin{figure}
\begin{overpic}{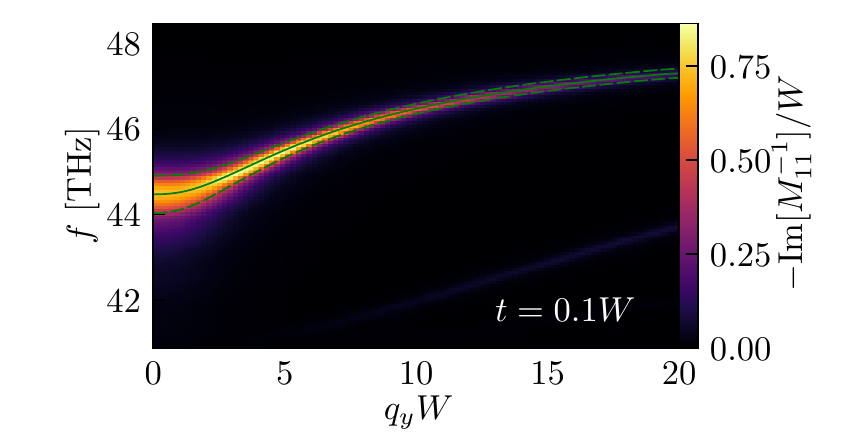}
\put(2,52){(a)}
\end{overpic}
\begin{overpic}{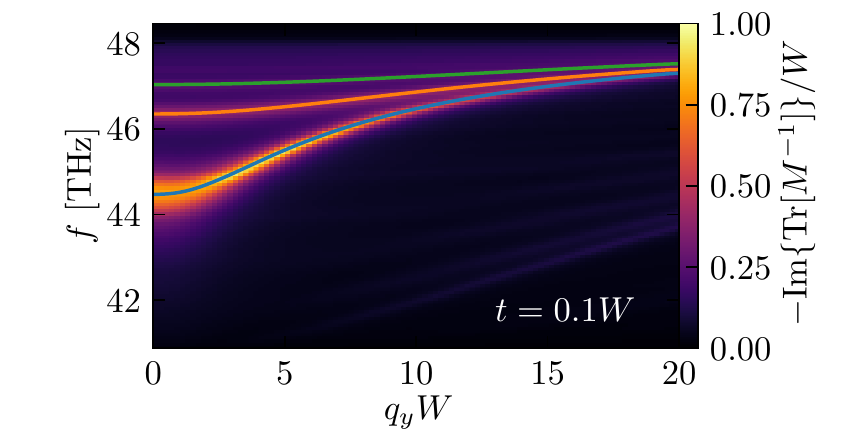}
\put(2,52){(b)}
\end{overpic}
\begin{overpic}{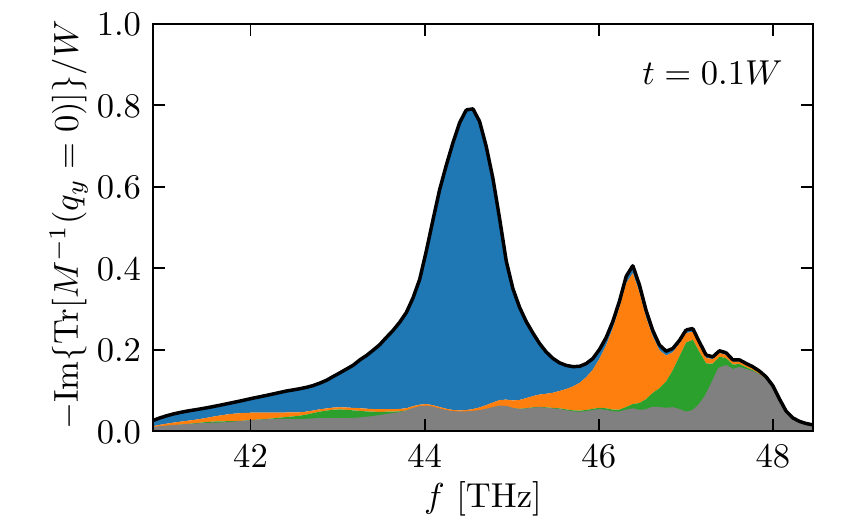}
\put(2,62){(c)}
\end{overpic}
\caption{\label{fig:loss} (a) First diagonal element of the matrix $M^{-1}$. Here $t = 0.1 W$.
(b) Trace of $M^{-1}$. The trace is done only on odd order modes.
(c) Line cut of (b) at $q_y = 0$, blue is the contribution from $M^{-1}_{ii}$ with $i =1$, orange $i=3$, green $i=5$.}
\end{figure}
\section{Green's functions of nanophotonics cavities}
\label{sect:cavities}
In this section we combine the Green's functions obtained in the previous section to build. 
We consider one-dimensional cavities, also dubbed ``line cavities'' or ``trench cavities'', and the limit of a very thick ($t'\gg W$) metal layer.
In what follows $\epsilon_{\parallel(\perp)}^{\pm}\equiv \epsilon_{\parallel(\perp)}(z=0^\pm)$.
Making use of Green's theorem (\ref{eq:green_theorem}) we can express the potential 
\begin{equation}\label{eq:green_theorem_1D}
\begin{split}
&\tilde{\phi}^{\pm}(x,z;q_y)  = \int dx'\int dz' \tilde{g}^\pm_0(x,z;x',z';q_y)\tilde{\rho}(x',z';q_y)\\
& \pm\frac{\epsilon_{\perp}^\pm}{4\pi}\int dx' \left.\partial_{z'}\tilde{g}^-_0(x,z;x',z';q_y)\right|_{z'=0} \tilde{\phi}^\pm(x',z=0;q_y),
\end{split}
\end{equation}
the corresponding $z$ component of the displacement field reads
\begin{equation}\label{eq:Dz_1D}
\begin{split}
&\tilde{D}_z^\pm(x,z=0,q_y)  =\\
& -\epsilon_\perp^\pm\int dx'\int dz' \left.\partial_z \tilde{g}^\pm_0(x,z;x',z';q_y)\right|_{z=0}\tilde{\rho}(x',z';q_y)\\
& \mp\frac{(\epsilon_\perp^\pm)^2}{4\pi}\int dx' \left.\partial_z\partial_{z'}\tilde{g}^-_0(x,z,x',z',q_y)\right|_{\substack{z'=0 \\z=0}} \tilde{\phi}^-(x',0,q_y).
\end{split}
\end{equation}
Since the potential has to be continuous at $z=0$ and vanishes at the metal we can expand it on the basis functions (\ref{eq:basis_functions_1D}) as $\phi^+(x,q_y,z=0) = \phi^-(x,q_y,z=0) = \sum_{n=1}^\infty \phi_n (q_y)\varphi_n(x)$, where the coefficients can be obtained as $\phi_n(q_y) = \int_{-W/2}^{W/2} dx \varphi_n(x) \phi^+(x,z=0;q_y)$. 
Moreover, since $D_z$ is continuous in the interval $[-W/2;W/2]$ we can write, for every integer $m>0$
\begin{equation}\label{eq:boundary_condition_1D}
\int dx \varphi_m(x)\left[\tilde{D}_z^+(x,z=0;q_y)-\tilde{D}_z^-(x,z=0;q_y)\right] = 0.
\end{equation}
Substitution of (\ref{eq:Dz_1D}) and straightforward algebra leads to
\begin{equation}\label{eq:response_matrix_1D}
\rho_m^+(q_y) + \rho_m^-(q_y)= \sum_{n=1}^\infty \left[M_{mn}^{-}(q_y)+M_{mn}^{+}(q_y)\right]\phi_n(q_y),
\end{equation}
where we defined
\begin{widetext}
\begin{align}
\rho_m^\pm(q_y) & \equiv \pm \epsilon_\perp^\pm \int dx \varphi_m(x)\int dx'\int dz'  \partial_z \left.\tilde{g}^\pm_0\left(x,z;x',z';q_y\right)\right|_{z=0}\tilde{\rho}(x',z';q_y), \label{eq:rho_m_def_1D}\\
M_{mn}^{\pm}(q_y)& \equiv - \frac{(\epsilon_\perp^\pm)^2}{4\pi} \int dx \varphi_m(x)\int dx' \varphi_n(x' ) \partial_{z}\partial_{z'} \left.\tilde{g}^\pm_0\left(x,z;x',z'\right)\right|_{\substack{z'=0 \\z=0}}.\label{eq:M_mn_def_1D}
\end{align}
Inverting the matrix $M_{nm}(q_y) \equiv M_{mn}^{-}(q_y)+M_{mn}^{+}(q_y)$ appearing in (\ref{eq:response_matrix_1D}) and substituting into (\ref{eq:green_theorem_1D}) we can put the full electrostatic Green's function of the system in the form
\begin{equation}\label{eq:green_function_1D}
\begin{split}
g(x,z;x',z';q_y) = & \Theta(z)\left\{\Theta(z')\left[ g_0^+(x,z;x',z';q_y) + \delta g^{++}(x,z;x',z';q_y)\right] + \Theta(-z')\delta g^{+-}(x,z;x',z';q_y) \right\}\\
 + & \Theta(-z)\left\{\Theta(-z')\left[ g_0^-(x,z;x',z';q_y) + \delta g^{--}(x,z;x',z';q_y)\right] + \Theta(z')\delta g^{-+}(x,z;x',z';q_y) \right\}.
\end{split}
\end{equation}
Here,
\begin{equation}\label{eq:delta_g_def_1D}
\begin{split}
\delta g^{\lambda\lambda'}(x,z;x',z';q_y) & = \frac{\lambda\lambda'\epsilon_\perp^\lambda \epsilon_\perp^{\lambda'}}{4\pi} \sum_{mn}  \left\{ \left[  \int dx''  \varphi_m(x'') \partial_{z''}g_0^\lambda(x,z;x'',z'';q_y)|_{z''=0} \right]
M_{nm}^{-1} (q_y) \right.\\
& \left.  \left[ \int dx'''  \varphi_n(x''') \partial_{z'''}g_0^{\lambda'}(x''',z''';x',z';q_y)|_{z'''=0} \right] \right\}\\
& = \frac{1}{4\pi} \sum_{nm}f_n^\lambda (x,z;q_y)M_{nm}^{-1}(q_y) f_m^{\lambda^\prime}(x',z';q_y)\\
& =\frac{1}{4\pi} \sum_{nm, j}f_n^\lambda (x,z;q_y)\frac{A_{nj}(q_y) [A^{-1}(q_y)]_{jm}}{\xi_j(q_y)} f_m^{\lambda^\prime}(x',z';q_y)\\
& = \frac{1}{4\pi} \sum_{j}\frac{\chi_j^\lambda(x,z;q_y)\Phi_j^{\lambda^\prime}(x',z';q_y)}{\xi_j(q_y)} .
\end{split}
\end{equation}
\begin{equation}
f^\lambda_m(x,z,q_y) = \lambda \epsilon_\perp^\lambda \int dx'\varphi_m(x') \partial_{z'} \left.g_0^\lambda (x,z;x',z';q_y)\right|_{z'=0}.
\end{equation}
\begin{equation}
M_{nm}(q_y) = A_{nj}(q_y)\xi_j(q_y) [A(q_y)]^{-1}_{jm}.
\end{equation}
\begin{equation}
\Phi_j(x,z;q_y) = \sum_m [A^{-1}(q_y) ]_{jm}f_m^\lambda (x,z;q_y)
\end{equation}
\begin{equation}
\chi_j(x,z;q_y) = \sum_m A_{mj}(qy) f_m^\lambda (x,z;q_y)
\end{equation}
\begin{equation}
f_m^+(x,z;q_y) =\int \frac{dq_x}{2\pi}e^{iq_x x} 8\pi   \hat{\varphi}_m(q_x)
\begin{cases}
\frac{\bar{\epsilon}e^{-q(z-t)}}{(\bar{\epsilon}+1)e^{q\eta t} + (\bar{\epsilon}-1)e^{-q\eta t}}\\
\frac{\bar{\epsilon}\cosh(q\eta(t-z))+\sinh(q\eta(t-z))}{(\bar{\epsilon}+1)e^{q\eta t} + (\bar{\epsilon}-1)e^{-q\eta t}}
\end{cases}
\end{equation}
\begin{equation}
f_m^-(x,z;q_y) = 4\pi \varphi_m(x) e^{\sqrt{q_y^2 +\frac{\pi^2 m^2}{W^2}}z}
\end{equation}

\begin{equation}\label{eq:delta_g_simplified}
\begin{split}
\delta g(x,z;x',z';q_y;\omega) = \frac{1}{4\pi} \sum_{j}\frac{\chi_j(x,z;q_y;\omega)\Phi_j(x',z';q_y;\omega)}{\xi_j(q_y;\omega)} .
\end{split}
\end{equation}

The matrix $M_{mn}^{+}(q_y)$, that depends only on the primitive Green's function for $z>0$ can be calculated as 
\begin{equation}\label{eq:M_plus_1D}
\begin{split}
M_{mn}^{+}(q_y)& = - \frac{(\epsilon_\perp^+)^2}{4\pi} \int_{-\infty}^\infty \frac{dq_x}{2\pi}\hat{ \varphi}_m^*(q_x) \hat{\varphi}_n(q_x)\partial_{z}\partial_{z'} \left.\hat{g}^+_0\left(z;z';\sqrt{q_x^2+q_y^2}\right)\right|_{\substack{z'=0 \\z=0}} =  \nonumber \\
&= \int_{-\infty}^\infty \frac{dq_x}{2\pi}\hat{ \varphi}_m^*(q_x) \hat{\varphi}_n(q_x) \bar{\epsilon}\sqrt{q_x^2+q_y^2} \frac{\bar{\epsilon}\tanh\left(\sqrt{q_x^2+q_y^2}\eta t\right)+1}{\tanh\left(\sqrt{q_x^2+q_y^2}\eta t\right)+ \bar{\epsilon}}= \nonumber\\
& = \frac{4\pi^2 \bar{\epsilon}i^{n-m}mn}{W^3}\int_{-\infty}^\infty\frac{dq_x }{2\pi}\frac{\sqrt{q_x^2+q_y^2} }{(q_x^2-\frac{m^2\pi^2}{W^2})(q_x^2-\frac{n^2\pi^2}{W^2})}\frac{\bar{\epsilon}\tanh\left(\sqrt{q_x^2+q_y^2}\eta t\right)+1}{\tanh\left(\sqrt{q_x^2+q_y^2}\eta t\right)+ \bar{\epsilon}}
\begin{cases}
1-\cos(q_xW)~m,n~\mbox{even}\\
1+\cos(q_xW)~m,n~\mbox{odd}\\
0 ~m,n~\mbox{opposite parity}.
\end{cases},
\end{split}
\end{equation}
where we made use of the Fourier transforms of the basis functions $\varphi_n(x)$
\begin{equation}\label{eq:basis_function_fourier}
\begin{split}
\hat{\varphi}_n(q_x)& = \int_{-\infty}^\infty dx e^{-ix \cdot q_x} \varphi_n(x) = \frac{2\sqrt{2}\pi n i^{n+1}}{W^{3/2}\left(q_x^2-\frac{n^2\pi^2}{W^2}\right)}\times
\begin{cases}
-\sin(q_xW/2) ~& n~\mbox{even}\\
\cos(q_xW/2) ~& n~\mbox{odd},\\
\end{cases}\\
& =  
\begin{cases}
i\frac{\sqrt{W}}{\sqrt{2}}\left[\sinc\left(\frac{q_xW}{2\pi}+\frac{n}{2}\right)-\sinc\left(\frac{q_xW}{2\pi}-\frac{n}{2}\right)\right] ~& n~\mbox{even}\\
\frac{\sqrt{W}}{\sqrt{2}}\left[\sinc\left(\frac{q_xW}{2\pi}+\frac{n}{2}\right)+\sinc\left(\frac{q_xW}{2\pi}-\frac{n}{2}\right)\right] ~& n~\mbox{odd},\\
\end{cases}\\
\end{split}
\end{equation}
and the last integral has to be done numerically.
Making use of the Green's function of the metallic cavity we obtain
\end{widetext}
\begin{equation}\label{eq:M_minus_1D_thick}
M_{mn}^{-}(q_y)  = \delta_{mn}\sqrt{q_y^2 + \frac{\pi^2n^2}{W^2}}.
\end{equation}
\section{Conclusions}
\label{sect:conclusions}
We can visualize the resonant features of the cavity response by plotting the interface loss function defined by $L(q_y, \omega) = -\im  [Tr(M^{-1}(q_y,\omega))]$.
This is shown in Fig.~\ref{fig:loss}, displaying a series of well-defined modes that disperse as a function of $q_y$.

To get a more quantitative information on the quality of these mode we fitted $L(q_y, \omega)$ at every value of $q_y$ with a Lorentzian function for each of the first three modes. From the fit parameter we can obtain the central frequency and the width of each of the modes as a function of $q_y$.
The results are reported in Fig.~\ref{fig:q_factor} in terms of the quality factor (ratio between the central frequency and the width) and the spatial quality factor or {\it finesse} that is given by the quality factor multiplied by the ratio of group velocity and phase velocity.

In summary, our calculation shows that nanophotonic cavities containing hyperbolic materials can have well-defined resonant modes even in the quasi-static regime. 

We stress that the Green's function contains all the information on the response of the cavity and allows studying the interaction of the cavities with other interesting systems, including quantum emitters, electronic systems and other resonant structures.

The present technique can be generalized to other simple but experimentally relevant geometries.
\begin{figure}[h!!]
\begin{overpic}{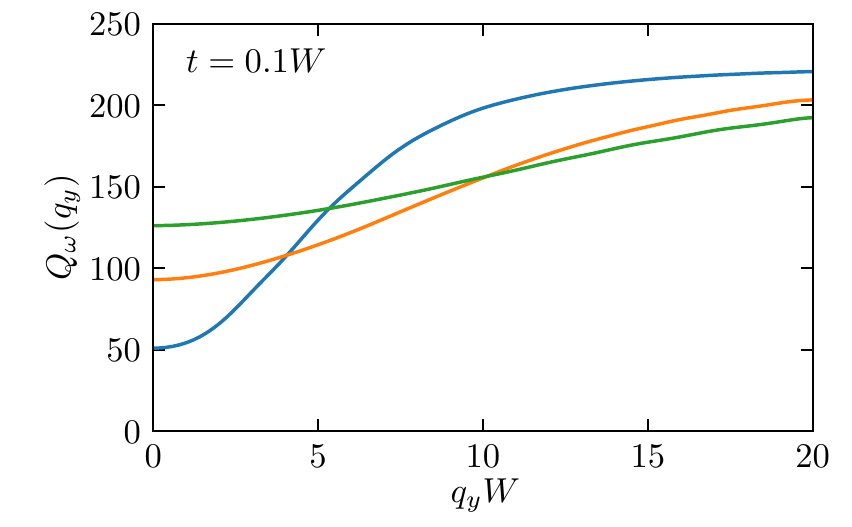}
\put(2,60){(a)}
\end{overpic}
\begin{overpic}{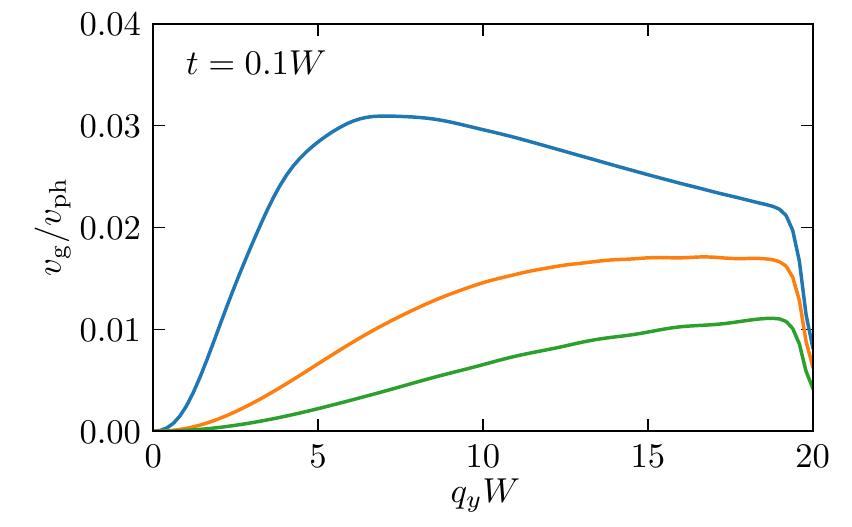}
\put(2,60){(b)}
\end{overpic}
\begin{overpic}{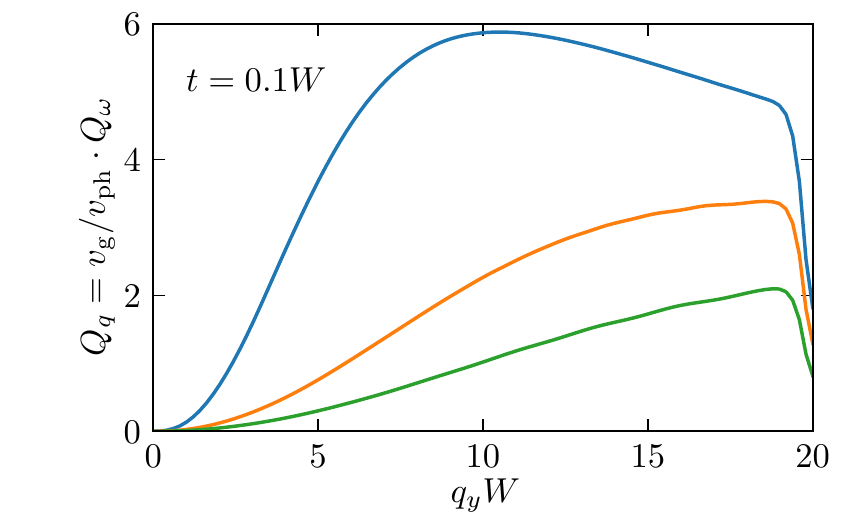}
\put(2,60){(c)}
\end{overpic}
\caption{\label{fig:q_factor} (a) Frequency quality factor $Q_\omega$ of the first three resonant modes obtained by fitting the resonances in Fig.\ref{fig:loss} with Lorentzian peaks. (b) Group velocity to phase velocity ratio for the same modes.
(c) Spatial quality factor $Q_q = v_{\rm g}/v_{\rm pg}Q_\omega$ for the same modes.
}
\end{figure}
\acknowledgments
We want to thank Prof. G.C. La Rocca and Dr. F. Lipparini for fruitful discussions.
I.T. acknowledges funding from the Spanish Ministry of Science, Innovation and Universities (MCIU) and State Research Agency (AEI) via the Juan de la Cierva fellowship Ref. FJC2018-037098-I.
F.H.L.K. acknowledges support by the ERC TOPONANOP under grant agreement n. 726001, the Government of Spain (FIS2016-81044; Severo Ochoa CEX2019-000910-S), Fundació Cellex, Fundació Mir-Puig, and Generalitat de Catalunya (CERCA, AGAUR, SGR 1656). Furthermore, the research leading to these results has received funding from the European Union’s Horizon 2020 under grant agreement no. 881603 (Graphene flagship Core3). H.H.S. acknowledges funding from the European Union’s Horizon 2020 programme under the Marie Skłodowska-Curie grant agreement Ref. 843830. 
\newpage
\onecolumngrid
\appendix
\section{Dyadic Green function}
\label{sect:dyadic}
The dyadic Green function relates the electric field generated at the position $\bm r $ to the external current density according to 
\begin{equation}
\bm E(\bm r,\omega) = \frac{i\omega}{c^2} \int d \bm r'  \bm G(\bm r,\bm r', \omega) \cdot  \bm J_{\rm ext}(\bm r',\omega),
\end{equation}
or, equivalently, the electric field generated by an external dipole $\bm p$ at the position $\bm r'$ to the dipole itself
\begin{equation}
\bm E(\bm r,\omega) = \frac{\omega^2}{c^2} \bm G(\bm r,\bm r', \omega) \cdot \bm p.
\end{equation}
Note that since it relates gauge-independent quantities the dyadic Green is gauge-independent.

The dyadic Green function solves the equation
\begin{equation}
\nabla \times \left\{ \bm \mu^{-1}(\bm r,\omega) \cdot \left[\nabla \times \bm G(\bm r,\bm r', \omega)\right]\right\} - \frac{\omega^2}{c^2} \bm \epsilon(\bm r,\omega) \cdot \bm G(\bm r,\bm r', \omega) = 4\pi\bm I \delta(\bm r-\bm r'),
\end{equation}
in absence of magnetic materials this simplifies to
\begin{equation}
\nabla \times \nabla\times \bm G(\bm r,\bm r', \omega) - \frac{\omega^2}{c^2} \bm \epsilon(\bm r,\omega) \cdot \bm G(\bm r,\bm r', \omega) = 4\pi \bm I \delta(\bm r-\bm r').
\end{equation}
In the static limit $\bm G(\bm r,\bm r', \omega)$ can be expressed in terms of the {\it electrostatic} Green function as
\begin{equation}
G_{ij}(\bm r,\bm r', \omega) = -\frac{c^2}{\omega^2}\partial_i \partial_j^\prime g(\bm r, \bm r'),
\end{equation}
where the electrostatic scalar Green function respects
\begin{equation}
-\nabla \cdot \left[{\bm \epsilon}(\bm r, \omega) \cdot \nabla g(\bm r, \bm r',\omega) \right] = 4\pi \delta(\bm r-\bm r').
\end{equation}
The total density of photonic states is proportional to the imaginary part of the trace of the dyadic Green function
\begin{equation}
\rho_{\rm ph} (\bm r, \omega) = \frac{\omega}{2\pi^2 c^2}\im\left[\tr \left[\bm G(\bm r,\bm r, \omega)\right]\right],
\end{equation}
in vacuum this equals
\begin{equation}
\rho_{\rm ph} ^{(0)}(\omega) = \frac{\omega^2}{\pi^2 c^3}.
\end{equation}
The density of states with a particular polarization orientation is instead
\begin{equation}
\rho_{\rm ph} (\bm r, \hat{\bm u},\omega) = \frac{3\omega}{2\pi^2 c^2}\im\left[\hat{\bm u} \cdot \bm G(\bm r,\bm r, \omega)\cdot\hat{\bm u}\right],
\end{equation}
Note that averaging this formula over the solid angle we obtain the total density of states.
The Purcell factor relative to the direction $\hat{\bm u}$ can be defined as
\begin{equation}
F_{\hat{\bm u}}(\bm r, \omega) \equiv \frac{\rho_{\rm ph} (\bm r, \hat{\bm u},\omega)}{\rho_{\rm ph} ^{(0)}(\omega)} = \frac{3c}{2\omega} \im\left[\hat{\bm u} \cdot \bm G(\bm r,\bm r, \omega)\cdot\hat{\bm u}\right]\approx -\frac{3c^3}{2\omega^3} \im\left[\hat{u}_i \hat{u}_j \partial_i\partial_j' g(\bm r,\bm r'\to \bm r, \omega)\right].
\end{equation}
\begin{equation}
\langle F\rangle(\bm r,\omega) \equiv \frac{\rho_{\rm ph} (\bm r, \hat{\bm u},\omega)}{\rho_{\rm ph} ^{(0)}(\omega)} = \frac{c}{2\omega} \im\left[\tr \left[\bm G(\bm r,\bm r, \omega)\right]\right]\approx -\frac{c^3}{2\omega^3} \im\left[ \partial_i\partial_i' g(\bm r,\bm r'\to \bm r, \omega)\right].
\end{equation}
\section{Uniqueness of the solution of the electrostatic problem}
\label{sect:uniqueness}
We consider the problem posed by (\ref{eq:poisson}) in a domain $\Omega$ with Dirichlet boundary conditions on a part of the boundary, i.e. $\phi(\bm r) = f_{\rm D}(\bm r)$ if $\bm r \in \partial \Omega_{\rm D}$ and Neumann boundary conditions on the remaining part ($\hat{\bm n}(\bm r) \cdot \bm \epsilon(\bm r) \cdot \nabla \phi(\bm r) = f_{\rm N}(\bm r)$ if $\bm r \in \partial \Omega_{\rm N}$) with $\partial\Omega = \Omega_{\rm D} \cup \partial \Omega_{\rm N}$ and $f_{\rm D}(\bm r), f_{\rm N}(\bm r)$ known functions.
The demonstration of the uniqueness of the solution of this problem closely parallels that of Laplace equation and proceeds by contradiction.

Let $\phi_1(\bm r)$ and $\phi_2(\bm r)$ be two continuous distinct solutions of the above problem.
Their difference $\psi(\bm r) = \phi_1(\bm r)-\phi_2(\bm r)$ is a solution of the homogeneous problem (\ref{eq:poisson}) with $\rho (\bm r) \equiv 0$, and $f_{\rm D}(\bm r), f_{\rm N}(\bm r)\equiv 0$.
The following integral therefore vanishes
\begin{equation}\label{eq:uniqueness}
0 = -\int_{\Omega} d\bm r \psi^*(\bm r) \nabla \cdot \left[\bm \epsilon (\bm r) \cdot \nabla \psi(\bm r)\right]
=-\oint_{\partial \Omega} d \bm s \cdot \psi^*(\bm r)\left[\bm \epsilon (\bm r) \cdot \nabla \psi(\bm r)\right]
+ \int_\Omega d \bm r \nabla \psi^*(\bm r) \cdot \bm \epsilon (\bm r) \cdot \nabla \psi(\bm r).
\end{equation}
The boundary term appearing in (\ref{eq:uniqueness}) vanishes because of homogeneous boundary conditions. 
By making use of the symmetry of $\bm \epsilon(\bm r)$ ($(\bm \epsilon +\bm \epsilon^\dagger)/2 = \bm \epsilon^{(1)}$, $(\bm \epsilon -\bm \epsilon^\dagger)/2 = i \bm \epsilon^{(2)}$) we can separate the real and imaginary part of the last integral that must vanish separately. This yields 
\begin{equation}\label{eq:uniqueness2}
\int_\Omega d \bm r \nabla \psi^*(\bm r) \cdot \bm \epsilon^{(i)} (\bm r) \cdot \nabla \psi(\bm r)=0.
\end{equation}
In non-hyperbolic dielectrics $\bm \epsilon^{(1)}$ is positive defined ensuring that $\nabla \psi(\bm r)$ must vanish everywhere. This means that $\psi(\bm r)$ is constant in each connected component of $\Omega$ and the two solutions are physically equivalent.

In the general case we can divide $\Omega$ into two non-overlapping domains $\Omega = \Omega_{\rm hyp} \cup \Omega_{\rm nor}$ such that $\bm \epsilon^{(1)}$ is positive defined in $\Omega_{\rm nor}$. 
We assume that $\bm \epsilon^{(2)}$ is semi positive-defined everywhere (absence of gain) and positive-defined in the region occupied by hyperbolic dielectrics.
Under this condition and applying (\ref{eq:uniqueness2}) to $\bm \epsilon^{(2)}$ we can prove that $\psi(\bm r)$ is piecewise constant in each connected component of $\Omega_{\rm hyp}$.
Applying again (\ref{eq:uniqueness2}) on the real part we can restrict the integral to $\Omega_{\rm nor}$, again we conclude that $\psi(\bm r)$ is piecewise constant in each connected component of $\Omega_{\rm nor}$.
By continuity $\psi(\bm r)$ is constant in each connected component of $\Omega$ and the two solutions are physically equivalent.
\section{Proof of Green's theorem in the presence of hyperbolic dielectrics}
\label{sect:green_proof}
Let's consider a finite domain $\Omega$. In this domain the dielectric function $\bm \epsilon(\bm r, \omega)$ is a symmetric complex matrix.
By product differentiation we obtain, for two generic functions $\varphi(\bm r)$ and $\psi(\bm r)$ 
\begin{equation}
\nabla \cdot [ \varphi(\bm r) \bm \epsilon(\bm r) \cdot \nabla \psi(\bm r)] = \varphi(\bm r) \nabla \cdot [ \bm \epsilon(\bm r) \cdot \nabla \psi(\bm r)]+ (\nabla \varphi(\bm r))\cdot \bm \epsilon(\bm r) \cdot (\nabla \psi(\bm r)),
\end{equation}
and, interchanging $\varphi$ and $\psi$
\begin{equation}
\nabla \cdot [ \psi(\bm r) \bm \epsilon(\bm r) \cdot \nabla \varphi(\bm r)] = \psi(\bm r) \nabla \cdot [ \bm \epsilon(\bm r) \cdot \nabla \varphi(\bm r)]+ (\nabla \psi(\bm r))\cdot \bm \epsilon(\bm r) \cdot (\nabla \varphi(\bm r)).
\end{equation}
subtracting the two and using divergence theorem we get the Green's identity
\begin{equation}\label{eq:green_identity}
\oint_{\partial \Omega}d \bm s \cdot [ \varphi(\bm r) \bm \epsilon(\bm r) \cdot \nabla \psi(\bm r) -  \psi(\bm r) \bm \epsilon(\bm r) \cdot \nabla \varphi(\bm r)] =\int_{\Omega} d\bm r \{\varphi(\bm r) \nabla \cdot [ \bm \epsilon(\bm r) \cdot \nabla \psi(\bm r)] - \psi(\bm r) \nabla \cdot [ \bm \epsilon(\bm r) \cdot \nabla \varphi(\bm r)]\}.
\end{equation}
Applying Green's identity to $\varphi(\bm r')= \phi(\bm r')$ and $\psi(\bm r') = g(\bm r,\bm r')$ yields
\begin{equation}
\begin{split}
\phi(\bm r, \omega) 
=  & \int_{\Omega} d\bm r' g(\bm r, \bm r', \omega)\rho_{\rm ext}(\bm r',\omega)\\
& + \frac{1}{4\pi} \oint_{\partial \Omega} ds' g(\bm r, \bm r') \left[\hat{\bm n}(\bm r')\cdot {\bm \epsilon}(\bm r', \omega) \cdot \nabla'  \phi(\bm r',\omega)\right]
 -\frac{1}{4\pi} \oint_{\partial \Omega} ds' \phi(\bm r', \omega) \left[\hat{\bm n}(\bm r') \cdot {\bm \epsilon}(\bm r', \omega) \cdot \nabla' g(\bm r, \bm r',\omega)\right].
\end{split}
\end{equation}
Making use of the Dirichlet boundary condition (\ref{eq:dirichlet_green}) leads to Eq.\ref{eq:green_theorem}.
Applying the Green's identity (\ref{eq:green_identity}) with $\varphi(\bm r)= g(\bm r_1,\bm r)$ and $\psi(\bm r) = g(\bm r,\bm r_2)$ proves the symmetry of the Green's function $g(\bm r,\bm r') = g(\bm r',\bm r)$ .

\onecolumngrid
\section{Derivatives of the green function}
\begin{align}
\partial_zg^+_{11}(z;z';q) 
= & -\frac{2\pi}{\epsilon_\perp}\sgn (z-z')e^{-q\eta|z-z'|}
+\frac{2\pi}{\epsilon_\perp}\frac{(\bar{\epsilon}+1)e^{-q\eta(z+z')}+(\bar{\epsilon}-1)e^{-2q\eta t}\left[e^{q\eta(z+z')}-2\sinh[q\eta(z-z')]\right]}{(\bar{\epsilon}+1)+(\bar{\epsilon}-1)e^{-2q\eta t}},\\
\partial_{z'}\partial_zg^+_{11}(z;z';q)  = 
& -\frac{2\pi \eta q}{\epsilon_\perp}e^{-q\eta|z-z'|}  +\frac{4\pi}{\epsilon_\perp}\delta(z-z')\nonumber\\
& -\frac{2\pi\eta q}{\epsilon_\perp}\frac{(\bar{\epsilon}+1)e^{-q\eta(z+z')}-(\bar{\epsilon}-1)e^{-2q\eta t}\left[e^{q\eta(z+z')}+2\cosh[q\eta(z-z')]\right]}{(\bar{\epsilon}+1)+(\bar{\epsilon}-1)e^{-2q\eta t}},\\
\partial_z g^+_{22}(z;z';q)= & -2\pi \sgn(z-z')e^{-q|z-z'|}
 +2\pi\frac{e^{q(2t-z-z')}\left[(\bar{\epsilon}-1)+(\bar{\epsilon}+1)e^{-2q\eta t}\right]}{(\bar{\epsilon}+1)+(\bar{\epsilon}-1)e^{-2q\eta t}},\\
\partial_{z'}\partial_z g^+_{22}(z;z';q)  = &
  -2\pi q e^{-q|z-z'|}
  +4\pi \delta(z-z')
-2\pi q\frac{e^{q(2t-z-z')}\left[(\bar{\epsilon}-1)+(\bar{\epsilon}+1)e^{-2q\eta t}\right]}{(\bar{\epsilon}+1)+(\bar{\epsilon}-1)e^{-2q\eta t}},
\end{align}
\twocolumngrid
\section{Optical constants of hexagonal boron nitride}
\label{sect:hbn}
The components of the dielectric tensor of hexagonal boron nitride (hBN) have the following frequency dependence~\cite{caldwell_naturecomm_2014} ($i=\parallel/ \perp$)
\begin{equation}\label{eq:hBN}
\epsilon_{i}(\omega) = \epsilon_{i}(\infty) + \frac{s_{i}\hbar^2 \omega^2_{i}}{\hbar^2 \omega^2_{i}-i\hbar^2 \gamma_{i}\omega-\hbar^2\omega^2}~,
\end{equation}
with parameters given in Table~\ref{table_parameters_hBN}. 
\begin{table}
\begin{tabular}{l | c c}
\, & $i=\parallel$  & $i=\perp$	\\
\hline
$s_{i}$	& 2.001 & 0.5262 \\
$\epsilon_{i}(\infty)$ & 4.9 & 2.95 \\	
$\hbar \omega_{i}~({\rm meV})$ & 168.6 & 94.2 \\
$\hbar \gamma_{i}~({\rm meV})$ & 0.87 & 0.25\\
\end{tabular}
\caption{The parameters entering the bulk hBN dielectric functions in Eq.~(\ref{eq:hBN}). These values have been extracted from Ref.~\onlinecite{caldwell_naturecomm_2014}.\label{table_parameters_hBN}}
\end{table}
\begin{figure}[h!!]
\centering
\begin{overpic}{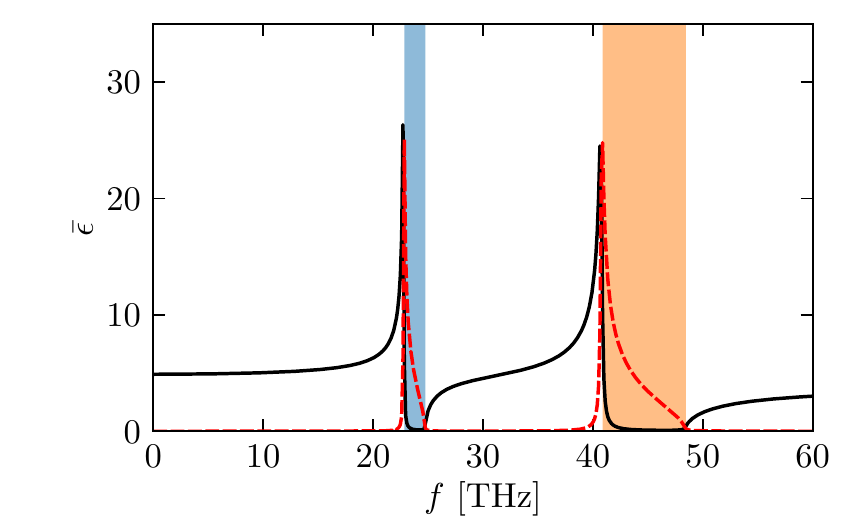}
\put(2,62){(a)}
\end{overpic}

\begin{overpic}{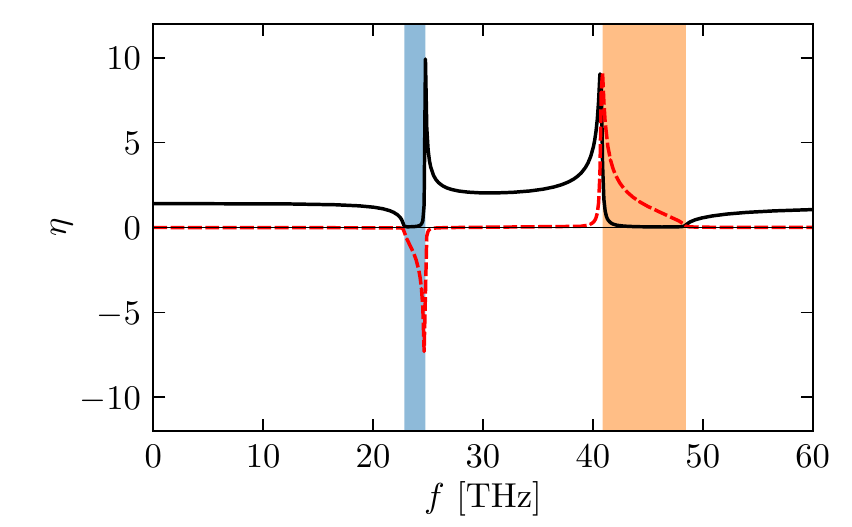}
\put(2,62){(b)}
\end{overpic}

\begin{overpic}{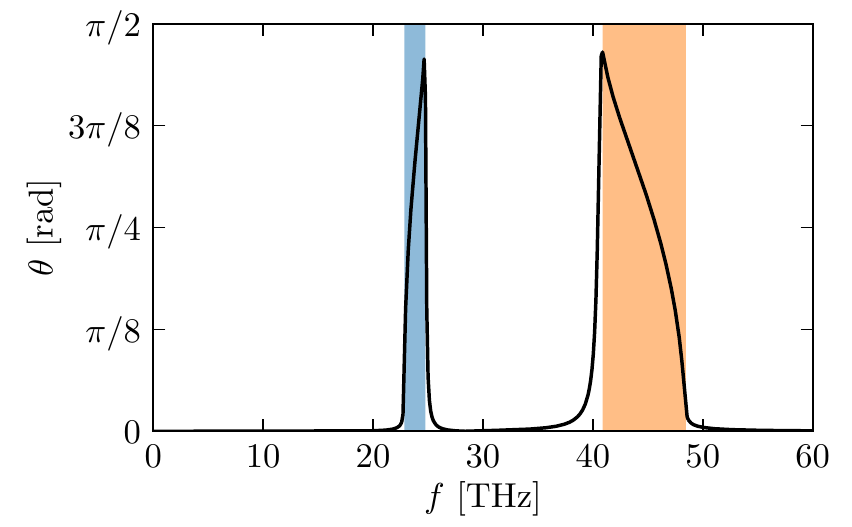}
\put(2,62){(c)}
\end{overpic}
\caption{\label{fig:hbn}
Optical constants of hexagonal Boron Nitride (h-BN). Black solid lines represent real parts, red dashed lines represent imaginary parts.}
\end{figure}
\twocolumngrid
\end{document}